# Lattice Boltzmann modeling of boiling heat transfer: The boiling curve and the effects of wettability

Q. Li [a, b], Q. J. Kang [b, *], M. M. Francois [c], Y. L. He [d], K. H. Luo [e]

[a] *School of Energy Science and Engineering, Central South University, Changsha 410083, China*

[b] *Computational Earth Science Group (EES-16), Los Alamos National Laboratory, Los Alamos, NM 87545, USA*

[c] *Fluid Dynamics and Solid Mechanics (T-3), Los Alamos National Laboratory, Los Alamos, NM 87545, USA*

[d] *MOE Key Laboratory of Thermal-Fluid Science and Engineering, Xi'an Jiaotong University, Xi'an 710049, China*

[e] *Department of Mechanical Engineering, University College London, London WC1E 7JE, UK*

*Corresponding author. Address: Computational Earth Science Group (EES-16), Mail Stop T003, Los Alamos National Laboratory, Los Alamos, NM 87545, USA. Tel.:+1 505 665 9663; Fax: +1 505 665 8737. Email: qkang@lanl.gov.

**Abstract**

A hybrid thermal lattice Boltzmann (LB) model is presented to simulate thermal multiphase flows with phase change based on an improved pseudopotential LB approach [Q. Li, K. H. Luo, and X. J. Li, Phys. Rev. E **87**, 053301 (2013)]. The present model does not suffer from the spurious term caused by the forcing-term effect, which was encountered in some previous thermal LB models for liquid-vapor phase change. Using the model, the liquid-vapor boiling process is simulated. The boiling curve together with the three boiling stages (nucleate boiling, transition boiling, and film boiling) is numerically reproduced in the LB community for the first time. The numerical results show that the basic features and the fundamental characteristics of boiling heat transfer are well captured, such as the severe fluctuation of transient heat flux in the transition boiling and the feature that the maximum heat transfer coefficient lies at a lower wall superheat than that of the maximum heat flux. Furthermore, the effects of the heating surface wettability on boiling heat transfer are investigated. It is found that an increase in contact angle promotes the onset of boiling but reduces the critical heat flux, and makes the boiling process enter into the film boiling regime at a lower wall superheat, which is consistent with the findings from experimental studies.

# 1. Introduction

Boiling is commonly observed in our daily life and is ubiquitous in nature as well as in many industrial applications [1-3]. Nucleate boiling, a well-known boiling phenomenon, has been recognized as one of the most effective heat transfer modes and used in a wide field of high-tech devices and systems such as nuclear reactors, heavy-vehicle engines, computer chips, and micro-electronic devices. Actually, boiling is an extremely complex and elusive process in which various physical components are involved and interrelated, such as the nucleation, growth, departure, and coalescence of vapor bubbles, the transport of latent heat, and the instability of liquid-vapor interfaces [4].

Owing to its intensive use and complexity, boiling has attracted extensive research efforts in the past decades and is still a subject of ongoing research activities in many groups all over the world. Nukiyaman is one of the pioneers in boiling research. He experimentally [5] measured the heat transmitted from metal to boiling water over a wide range of wall superheats and qualitatively established the pool boiling curve, which is an important corner-stone for boiling research and helps to distinguish different regimes in pool boiling [6]. Today, it is widely recognized that the pool boiling with controlled surface temperature can be divided into three distinct regimes [1, 3, 6]: nucleate boiling, transition boiling, and film boiling.

*Nucleate boiling* is a boiling mode that takes place when the temperature of the heating surface is higher than the saturated fluid temperature by a certain amount but the heat flux is below the critical heat flux. *Film boiling* appears when the surface temperature is increased above a threshold temperature (the so-called Leidenfrost temperature). In film boiling, the heating surface is completely covered with a continuous vapor film and the liquid does not contact the heating surface. *Transition boiling*, in which a large portion of the heating surface will be covered by vapor, occurs at surface temperatures between the maximum attainable in nucleate boiling and the minimum attainable in film boiling. This type of boiling is very unstable as it is accompanied by a reduction in the heat flux with an increase in the wall superheat [3].

With the rapid development of computer technology, numerical simulations of boiling phenomena gradually play an important role in investigating the mechanism and the heat transfer characteristics of boiling [6]. The first attempt was made by Son and Dhir [7, 8], who studied the evolution of the liquid-vapor interface during saturated film boiling with a level set method [8]. Meanwhile, Juric and Tryggvason [9] extended the front tracking method to simulate horizontal film boiling by adding a source term to the continuity equation. Subsequently, Welch and Wilson [10] proposed a volume of fluid based method to simulate horizontal film boiling. Since then, a lot of numerical studies have been conducted to investigate boiling phenomena within the framework of the traditional numerical methods, in which the liquid-vapor interface is tracked as part of the solution of the mass, momentum, and energy conservation equations. Detailed review of these studies can be found in Refs. [3, 6, 11, 12].

In recent years, the lattice Boltzmann (LB) method, which is a mesoscopic approach based on minimal lattice formulations of the kinetic Boltzmann equation [13, 14], has also been applied to simulate liquid-vapor phase change. Generally, the existing thermal LB models for liquid-vapor phase change can be classified into two categories. The first category is based on the phase-field LB method. The thermal LB models developed by Dong *et al.* [15], Safari *et al.* [16], and Sun *et al.* [17] fall into this category. In these models, the interface-capturing equation (the Cahn-Hilliard equation) is solved to capture the liquid-vapor interface and a source term is incorporated into the continuity equation or the Cahn-Hilliard equation to define the phase change, which means that in these models the rate of the liquid-vapor phase change is an artificial input. In addition, the models in Refs. [15, 17] are based on the phase-field LB model proposed by Zheng *et al.* [18], which has been demonstrated [19] to be a binary model and give the same results for the cases with the same average number density, e.g., for the case $n_A = 1000$ and $n_B = 1$ with the case $n_A = 501$ and $n_B = 500$ [19].

The second category is based on the pseudopotential LB approach [20-22], which is a very popular multiphase approach in the LB community. The most distinct feature of the pseudopotential LB method is that the phase separation between different phases is achieved via an interparticle potential.

As a result, the liquid-vapor interfaces can naturally arise, deform, and migrate without using the interface-tracking or interface-capturing techniques, which is usually required in the traditional numerical methods and in the above-mentioned phase-filed LB method. The first thermal LB model based on the pseudopotential LB approach may be attributed to Zhang and Chen [23], who devised a new force expression and modeled the nucleate boiling. Later, Házi and Márkus [24-26], Biferale *et al*. [27, 28], and Cheng *et al*. [29, 30] also proposed thermal LB models based on the pseudopotential LB approach. These models share the feature that a thermal LB equation with a temperature distribution function is solved to mimic the macroscopic temperature equation. The heat conduction term was split into two parts [26, 28]: one was recovered by the thermal LB equation whereas the other part was calculated with finite-difference schemes and then incorporated into the thermal LB equation via source terms.

In fact, for thermal LB equations, the forcing term of the system will introduce a spurious term into the macroscopic temperature equation. The numerical errors caused by this spurious term can be neglected in most cases. However, it was recently found that [31] such a forcing-term effect will lead to significant errors in the pseudopotential LB modeling of thermal flows. Furthermore, for temperature-based thermal LB equations, another error term proportional to $\nabla \cdot (T\nabla\rho/\rho)$ will also exist in the recovered temperature equation, which can be found following the derivations in Ref. [32] (see Eq. (50) therein) and the derivations in Ref. [33]. This term will also yield considerable errors for multiphase flows (variable density).

In the present work we therefore adopt the hybrid thermal LB method, which still uses a LB model to simulate the fluid flow but solves the temperature field with a traditional numerical scheme such as the finite-difference or finite-volume scheme. Hence it is free of the above-mentioned problems. In addition, the original pseudopotential LB approach usually suffers from the problem of thermodynamic inconsistency, i.e., the predicted liquid-vapor coexistence densities are inconsistent with the results given by the Maxwell construction [34]. We recently found that the thermodynamic inconsistency can be eliminated through adjusting the mechanical stability condition via an improved forcing scheme [35, 36]. Hence in this study we use the improved pseudopotential LB model [36] to simulate the fluid flow. The temperature equation is solved by a finite-difference scheme. The coupling between the pseudopotential LB model and the finite-difference scheme is established via a non-ideal equation of state. The rest of the present paper is organized as follows. The hybrid thermal LB model is formulated in Sec. 2. Numerical simulation of boiling heat transfer will be presented in Sec. 3. Analysis and discussion will also be made. Finally, a brief summary will be given in Sec. 4.

## 2. The hybrid thermal LB model

The pseudopotential LB approach was developed by Shan and Chen in 1993 [20]. They proposed an interparticle potential, which is now widely called pseudopotential, to mimic the fluid interactions. In Shan and Chen's original pseudopotential LB model, the Bhatnagar-Gross-Krook (BGK) collision operator [37] was employed. In recent years, it has been found that the Multi-Relaxation-Time (MRT) collision operator [38, 39] performs much better than the BGK operator in terms of numerical stability [40, 41]. The LB equation with a MRT collision operator can be given as follows:

$$f_\alpha \left( \boldsymbol{x} + \boldsymbol{e}_\alpha \delta_t, t + \delta_t \right) = f_\alpha \left( \boldsymbol{x}, t \right) - \left( \mathbf{M}^{-1} \mathbf{\Lambda} \mathbf{M} \right)_{\alpha\beta} \left( f_\beta - f_\beta^{eq} \right) + \delta_t F'_\alpha, \quad (1)$$

where $f_\alpha$ is the density distribution function, $f_\alpha^{eq}$ is the equilibrium distribution, $\boldsymbol{x}$ is the spatial position, $\boldsymbol{e}_\alpha$ is the discrete velocity along the $\alpha$ th lattice direction, $\delta_t$ is the time step, $F'_\alpha$ is the forcing term in the velocity space, $\mathbf{M}$ is an orthogonal transformation matrix, and $\mathbf{\Lambda}$ is a diagonal Matrix. Using the transformation matrix $\mathbf{M}$, the right-hand side of Eq. (1) can be rewritten as [41]

$$\mathbf{m}^* = \mathbf{m} - \mathbf{\Lambda} \left( \mathbf{m} - \mathbf{m}^{eq} \right) + \delta_t \left( \mathbf{I} - \frac{\mathbf{\Lambda}}{2} \right) \mathbf{S}, \quad (2)$$

where $\mathbf{m} = \mathbf{M} \mathbf{f}$, $\mathbf{m}^{eq} = \mathbf{M} \mathbf{f}^{eq}$, $\mathbf{I}$ is the unit tensor, and $\mathbf{S}$ is the forcing term in the moment space with $\left( \mathbf{I} - 0.5 \mathbf{\Lambda} \right) \mathbf{S} = \mathbf{M} \mathbf{F}'$. Then the MRT LB equation can be formulated as

$$f_\alpha \left( \boldsymbol{x} + \boldsymbol{e}_\alpha \delta_t, t + \delta_t \right) = f_\alpha^* \left( \boldsymbol{x}, t \right), \quad (3)$$

where $\mathbf{f}^* = \mathbf{M}^{-1}\mathbf{m}^*$. In the LB method, Eq. (2) and Eq. (3) are usually called "collision" and "streaming", respectively. For the D2Q9 lattice, the diagonal Matrix $\Lambda$, which includes the relaxation times, is given by

$$\Lambda = \mathrm{diag}\left(\tau_\rho^{-1}, \tau_e^{-1}, \tau_\varsigma^{-1}, \tau_j^{-1}, \tau_q^{-1}, \tau_j^{-1}, \tau_q^{-1}, \tau_\upsilon^{-1}, \tau_\upsilon^{-1}\right). \tag{4}$$

The meaning of these relaxation times and the transformation matrix $\mathbf{M}$ can be found in Ref. [38]. Through $\mathbf{m}^{eq} = \mathbf{M}\mathbf{f}^{eq}$, the equilibria $\mathbf{m}^{eq}$ can be obtained as follows:

$$\mathbf{m}^{eq} = \rho\left(1, -2+3|\mathbf{v}|^2, 1-3|\mathbf{v}|^2, v_x, -v_x, v_y, -v_y, v_x^2 - v_y^2, v_x v_y\right)^\mathrm{T}, \tag{5}$$

where $\mathbf{v}$ is the macroscopic velocity and $|\mathbf{v}| = \sqrt{v_x^2 + v_y^2}$. The macroscopic density and velocity are calculated via

$$\rho = \sum_\alpha f_\alpha, \quad \rho\mathbf{v} = \sum_\alpha \mathbf{e}_\alpha f_\alpha + \frac{\delta_t}{2}\mathbf{F}, \tag{6}$$

where $\mathbf{F} = (F_x, F_y)$ is the total force acting on the system. The most important force in the pseudopotential LB model is the intermolecular interaction force, which is given by [42]

$$\mathbf{F}_m = -G\psi(\mathbf{x})\sum_\alpha w_\alpha \psi(\mathbf{x} + \mathbf{e}_\alpha)\mathbf{e}_\alpha, \tag{7}$$

where $\psi$ is the pseudopotential, $G$ is the interaction strength, and $w_\alpha$ are the weights. In this work, the pseudopotential $\psi$ is taken as $\psi(\mathbf{x}) = \sqrt{2(p_{\mathrm{EOS}} - \rho c_s^2)/Gc^2}$ [43], in which $p_{\mathrm{EOS}}$ is the prescribed non-ideal equation of state. In order to achieve thermodynamic consistency, an improved forcing scheme which was presented in our recent work [36] is utilized for the forcing term $\mathbf{S}$:

$$\mathbf{S} = \begin{bmatrix} 0 \\ 6\mathbf{v}\cdot\mathbf{F} + \dfrac{\sigma|\mathbf{F}_m|^2}{\psi^2 \delta_t (\tau_e - 0.5)} \\ -6\mathbf{v}\cdot\mathbf{F} - \dfrac{\sigma|\mathbf{F}_m|^2}{\psi^2 \delta_t (\tau_\varsigma - 0.5)} \\ F_x \\ -F_x \\ F_y \\ -F_y \\ 2(v_x F_x - v_y F_y) \\ (v_x F_y + v_y F_x) \end{bmatrix}, \tag{8}$$

where $\sigma$ is an parameter used to tune the mechanical stability condition, $|\mathbf{F}_m|^2 = \left( F_{m,x}^2 + F_{m,y}^2 \right)$, and $\mathbf{F}$ is the total force, which also contains the buoyant force given by

$$\mathbf{F}_b = (\rho - \rho_{ave})\mathbf{g}, \tag{9}$$

where $\rho_{ave}$ is the average density in the computational domain and $\mathbf{g} = (0, -g)$ is the gravitational acceleration. The validation and verification of the improved forcing scheme can be found in Ref. [36].

Through the Chapman-Enskog analysis of the LB equation, the following macroscopic equations can be obtained at the Navier-Stokes level:

$$\partial_t \rho + \nabla \cdot (\rho \mathbf{v}) = 0, \tag{10}$$

$$\partial_t (\rho \mathbf{v}) + \nabla \cdot (\rho \mathbf{v}\mathbf{v}) = -\nabla \cdot \mathbf{P} + \nabla \cdot \mathbf{\Pi} + \mathbf{F}_b, \tag{11}$$

where $\mathbf{\Pi}$ is the viscous stress tensor and $\mathbf{P}$ is the pressure tensor, which can be attained according to the intermolecular interaction force Eq. (7) and is given by

$$\mathbf{P} = \left( p_{EOS} + \frac{G^2 c^4}{6} \sigma |\nabla \psi|^2 + \frac{G c^4}{12} \psi \nabla^2 \psi \right) \mathbf{I} + \frac{G c^4}{6} \psi \nabla \nabla \psi, \tag{12}$$

where $c$ is the lattice constant. The last term in Eq. (12) is related to the surface tension.

Now attention turns to solving the temperature equation. For the diffuse interface modeling of multiphase flows, Anderson *et al*. [44] have summarized the local balance law for entropy, which is given by (neglecting the viscous heat dissipation)

$$\rho T \frac{Ds}{Dt} = \nabla \cdot (\lambda \nabla T), \tag{13}$$

where $s$ is the entropy, $\lambda$ is the thermal conductivity, and $D(\bullet)/Dt = \partial_t (\bullet) + \mathbf{v} \cdot \nabla (\bullet)$. The temperature equation can be derived from Eq. (13) using the following thermodynamic relation [25]:

$$T ds = c_v dT + T \left( \frac{\partial p_{EOS}}{\partial T} \right)_\rho d\left( \frac{1}{\rho} \right), \tag{14}$$

where $c_v$ is the specific heat at constant volume. Then the temperature equation is given by

$$\rho c_v \frac{DT}{Dt} = \nabla \cdot (\lambda \nabla T) - T \left( \frac{\partial p_{EOS}}{\partial T} \right)_\rho \nabla \cdot \mathbf{v}, \tag{15}$$

Obviously, for ideal gases ($p_{EOS} = \rho RT$), the last term yields $p_{EOS}\nabla\cdot\mathbf{v}$. The above equation can also be rewritten as

$$\partial_t T + \mathbf{v}\cdot\nabla T = \frac{1}{\rho c_v}\nabla\cdot(\lambda\nabla T) - \frac{T}{\rho c_v}\left(\frac{\partial p_{EOS}}{\partial T}\right)_\rho \nabla\cdot\mathbf{v} \tag{16}$$

In Ref. [26], Házi et al. rewrote Eq. (16) as follows

$$\underline{\partial_t T + \nabla\cdot(\mathbf{v}T) = \nabla\cdot(k\nabla T)} + \frac{1}{\rho c_v}\nabla\cdot(\lambda\nabla T) - \nabla\cdot(k\nabla T) + \left[T - \frac{T}{\rho c_v}\left(\frac{\partial p_{EOS}}{\partial T}\right)_\rho\right]\nabla\cdot\mathbf{v}, \tag{17}$$

The underlined terms were claimed to be solved by a thermal LB equation and $k$ is determined by the relaxation time in the thermal LB equation. The other terms were calculated with finite-difference schemes and then incorporated into the thermal LB equation via source terms. The problems of Házi et al.'s model have been mentioned in the introduction, i.e., suffering from the spurious term caused by the forcing-term effect and the error term proportional to $\nabla\cdot(T\nabla\rho/\rho)$. Moreover, as can be seen in Eq. (17), all the source terms still need to be calculated with finite-difference schemes.

We therefore directly use the finite-difference method to solve Eq. (16), which is equivalent to

$$\partial_t T = -\mathbf{v}\cdot\nabla T + \frac{1}{\rho c_v}\nabla\cdot(\lambda\nabla T) - \frac{T}{\rho c_v}\left(\frac{\partial p_{EOS}}{\partial T}\right)_\rho \nabla\cdot\mathbf{v}. \tag{18}$$

For simplicity, the right-hand side of Eq. (18) is denoted by $K(T)$. The classical fourth-order Runge-Kutta scheme [45, 46] is adopted for time discretization:

$$T^{t+\delta_t} = T^t + \frac{\delta_t}{6}(h_1 + 2h_2 + 2h_3 + h_4), \tag{19}$$

where $h_1$, $h_2$, $h_3$, and $h_4$ are given by, respectively

$$h_1 = K(T^t), \quad h_2 = K\left(T^t + \frac{\delta_t}{2}h_1\right), \quad h_3 = K\left(T^t + \frac{\delta_t}{2}h_2\right), \quad h_4 = K(T^t + \delta_t h_3). \tag{20}$$

The quantities $\rho$ and $\mathbf{v}$ are given by the pseudopotential LB model. Hence in above computations, $\rho$ and $\mathbf{v}$ are at the $t$ level. For the spatial discretization, the isotropic central schemes proposed by Lee and Lin [47] are employed to evaluate the first-order derivative and the Laplacian.

In summary, Eqs. (2), (3), (6), and (8) together with Eqs. (19) and (20) constitute the hybrid thermal LB model. The pseudopotential LB model and the finite-difference solver of the temperature equation is coupled using the non-ideal equation of state $p_{\text{EOS}}$, which is incorporated into the pseudopotential LB model via $\psi(\mathbf{x}) = \sqrt{2(p_{\text{EOS}} - \rho c_s^2)/Gc^2}$. In this work, the Peng-Robinson (P-R) equation of state is used [43]

$$p_{\text{EOS}} = \frac{\rho RT}{1-b\rho} - \frac{a\varphi(T)\rho^2}{1+2b\rho - b^2\rho^2}, \tag{21}$$

where $\varphi(T) = \left[1 + \left(0.37464 + 1.54226\omega - 0.26992\omega^2\right)\left(1 - \sqrt{T/T_c}\right)\right]^2$, $a = 0.45724 R^2 T_c^2 / p_c$, and $b = 0.0778 RT_c / p_c$. The parameter $\omega$ is the acentric factor. Yuan and Schaefer [43] have shown that, when $\omega = 0.344$, the coexistence curve given by the P-R equation of state is close to the experiment data of saturated water/steam. According to Ref. [43] and the relationship between $a$ and the interface thickness [36], in the present work we choose $a = 3/49$, $b = 2/21$, and $R = 1$. For such a choice, the parameter $\sigma$ in Eq. (8) can be set to $1.2$ for the sake of achieving thermodynamic consistency.

## 3. Numerical results and analyses

As previously mentioned, boiling is an extremely complex process. Although several studies have been conducted within the LB framework for simulating liquid-vapor phase change, the LB modeling of boiling phenomena is still in its infancy. Up to now, the modeling of the complete three boiling stages as well as the boiling curve has not been realized. In this section, using the proposed hybrid thermal LB model, we make an attempt to simulate boiling heat transfer with a focus on the three different boiling regimes and the boiling curve.

### 3.1 The simulation setup

Numerical simulations are carried out in the two-dimensional $x-y$ plane with fluids confined in the computational domain $0 \leq x \leq L_x$ and $0 \leq y \leq L_y$ with $L_x \times L_y = 600 \times 150$. The studied problem is similar to that in Ref. [23]. Namely, two solid walls are defined at $y = 0$ and $y = L_y$, and the periodic boundary condition is applied in the $x$ direction to close the system. The initial setting of the computational domain is a liquid ($0 \leq y < 90$) below its vapor ($90 \leq y < L_y$), and the temperature in the domain is the corresponding coexistence temperature $T_s$. The bottom wall is a heating surface with a high temperature $T_b$ while the temperature at the top wall is maintained at $T_s$.

Our simulations start from the equilibrium state of two-phase coexistence at the temperature $T_s = 0.86 T_c$, which corresponds to the coexistence densities $\rho_L \approx 6.5$ and $\rho_V \approx 0.38$. The Zou-He non-slip boundary condition [48] is employed at the solid walls. The kinematic viscosities of the liquid and vapor phases are taken as $\upsilon_L = 0.1$ and $\upsilon_V = 0.5/3$, respectively, with the dynamic viscosity ratio being $\mu_L/\mu_V = (\rho_L \upsilon_L)/(\rho_V \upsilon_V) \approx 10$. Following Refs. [23, 24], a constant $c_v$ is adopted in the present study and is set to 6.0. For simplicity, the thermal conductivity $\lambda = \rho c_v \chi$, where $\chi$ is the thermal diffusivity, is chosen to be proportional to the density $\rho$ with $c_v \chi = 0.028$. Then $\lambda_L/\lambda_V = \rho_L/\rho_V \approx 17$. The gravitational acceleration is taken as $g = 3 \times 10^{-5}$. Note that the values of the above quantities are based on the lattice unit with the lattice constant $c = \delta_x/\delta_t = 1$ and $\delta_x = \delta_t = 1$. Using the LB method, numerical simulations are usually conducted in lattice units. The conversion between the lattice units and the physical units can be found elsewhere [49-51].

**3.2 The three boiling stages and the boiling curve**

The intermolecular (fluid-fluid) interaction force Eq. (7) is not applied at the solid walls. With this treatment, the equilibrium contact angle (defined in the liquid phase) is around $44.5°$. Note that, if the fluid-fluid interaction force is employed at the solid walls with a ghost fluid layer being placed below the solid walls, the corresponding contact angle is about $90°$. The contact angle of the heating surface can be adjusted via a fluid-solid interaction force [33]. Here we firstly consider the cases without the fluid-solid interaction force. Following the treatment in Ref. [23], small fluctuations are added to enhance the bubble formation. The wall superheat is defined as $\Delta T = T_b - T_s$, which is also given in lattice unit and is changed by tuning the bottom-wall temperature $T_b$.

Figure 1 shows the time evolution of the density contours at $\Delta T = 0.0137$ (lattice unit). From Fig. 1(a) ($t = 10000\delta_t$) we can see that, owing to the high temperature, low density fluids are produced near the bottom wall, which lead to a vapor film in the center and some small vapor bubbles on the sides. Subsequently, as can be seen in Fig. 1(b), the vapor film will grow and yield satellite vapor bubbles in the rims due to the instability of the interface. Besides, some small bubbles will coalesce with each other to form larger size ones, which can be seen by comparing Fig. 1(b) with Fig. 1(a), and Fig. 1(c) with Fig. 1(b). Later, the vapor bubbles will gradually depart from the bottom wall and move upward because of the buoyancy force. Then the rising vapor bubbles will merge into the vapor region at the top as can be seen at $t = 30000\delta_t$. Meanwhile, new vapor bubbles appear at the bottom wall and begin to grow. It is obvious that the boiling at $\Delta T = 0.0137$ is in the nucleate boiling regime.

Now we consider the cases with higher wall superheats. To see the differences between these cases more clearly, the wall superheat $\Delta T$ is increased at $t = 20000\delta_t$ from the previous wall superheat $\Delta T = 0.0137$ to the higher wall superheats. Figure 2 shows the results of the case $\Delta T = 0.0165$. From the figure we can see that, when the wall superheat is increased to $0.0165$, more nucleation sites are activated near the bottom wall ($t = 22000\delta_t$), from which many vapor bubbles will be generated (as can be seen at $t = 26000\delta_t$). Furthermore, it can be clearly observed that the number density of the growing and rising bubbles is increased and the interaction between the vapor bubbles is stronger than that in the case of $\Delta T = 0.0137$. According to the results in Fig. 2, we believe that the boiling at $\Delta T = 0.0165$ is still in the nucleate boiling regime but with higher heat transfer rate.

With more and more vapor bubbles rising from the bottom, the pressure at the top will increase due to the confinement of the top wall, which can be seen in Fig. 3, where the variation of the pressure at the central point of the top wall is shown for the previous case ($\Delta T = 0.0137$) and the present case. When a rising bubble merges into the vapor region at the top, it should have the same pressure as that in the top vapor region. Hence the rising bubble will shrink to some extent so as to increase its pressure (the Young-Laplace law $\Delta p = \vartheta/r$), which can be observed in Figs. 2(a), 2(b), and 2(c) for the bubbles near the top. The shrinkage of rising bubbles can also be found in Ref. [52] (see Fig. 12 therein), where direct numerical simulations of nucleate boiling flows in a cavity were conducted.

Figure 4 displays the results of the case that the wall superheat is raised to $0.032$ at $t = 20000\delta_t$. As can be seen in Fig. 4, the most apparent feature of this case is that a great portion of the heating surface is covered by vapor patches. Since vapor is less capable of conducting heat, these vapor patches will essentially insulate the bulk liquid from the heating surface and lead to a decrease in the heat transfer rate. According to the results in Fig. 4, the boiling at $\Delta T = 0.032$ may be in the transition boiling regime.

Figure 5 gives the results of the case $\Delta T = 0.035$. We can see that a further increase in the wall superheat results in the whole heating surface being covered with a continuous vapor film, which completely separates the bulk liquid from the heating surface. Therefore the heat is transferred from the heating surface to the bulk liquid by the conduction and convection through the vapor film with the evaporation at the liquid-vapor interface. When the liquid approaches the heating wall, the evaporation at the liquid-vapor interface will prevent the liquid from contacting the wall and will provide vapor to form rising bubbles, which can be seen from Figs. 5(c), 5(d), and 5(e). Clearly, the boiling at $\Delta T = 0.035$ is in the film boiling regime, which is of great interest to certain applications such as quenching of steel or spray cooling of very hot surfaces.

The transient heat flux defined by $q(t) = \int_0^{L_x} q_w(x) dx / L_x$ is plotted in Fig. 6 for the above three cases, where $q_w(x) = -\lambda (\partial T/\partial y)\big|_{y=0}$ is the local heat flux at the bottom wall. From the figure we can see that, due to the low thermal conductivity of the vapor film, the heat flux in the case $\Delta T = 0.035$ is very small as compared with the other two cases. More importantly, it can be seen that the heat flux fluctuates significantly and frequently in the case $\Delta T = 0.032$, while the heat flux in the cases $\Delta T = 0.0165$ and $\Delta T = 0.035$ is much more stable. Such a feature further indicates that the boiling at $\Delta T = 0.032$ may be in the transition boiling regime. In fact, the transition boiling can be viewed as an unstable combination of the nucleate boiling and the film boiling, each of which alternately exists at any given location on the heating surface [53-54]. The variation of the transient heat flux is primarily a result of the change in the fractions of these two boiling modes on the whole heating surface.

In order to show the boiling regimes more clearly, the boiling curve is given in Fig. 7, which is established by averaging the transient heat flux over time. For completeness, some other cases have also been studied so as to complete the boiling curve. At the cases $\Delta T = 0.008$ and $0.01$, the wall superheat is insufficient to support bubble formation and growth. Hence no boiling occurs and the heat flux is very low as can be seen in Fig. 7. With the further increase of the wall superheat, boiling will occur and the heat flux will increase rapidly until reaching its maximum value (the critical heat flux) around $\Delta T = 0.021 \sim 0.025$. After that the heat flux decreases dramatically as the bulk liquid will be insulated from the heating wall by more and more vapor. The points "A" and "B" in the boiling curve denotes the cases $\Delta T = 0.0165$ and $0.032$, respectively. Clearly, the point "A" is in the high-heat-flux nucleate boiling regime [6] whereas the point "B" is in the transition boiling regime.

The corresponding heat transfer coefficient $h_c = \overline{q}/\Delta T$, which is the ratio of the average heat flux to the thermodynamic driving force for boiling (i.e., the wall superheat), is displayed in Fig. 8. The heat transfer coefficient can be used to describe the thermodynamic efficiency of the boiling exchange. By comparing Fig. 8 with Fig. 7, we can observe an important feature of boiling heat transfer, namely the maximum heat transfer coefficient lies at a lower wall superheat (point "A") than that of the maximum heat flux. Actually, in the nucleate boiling regime, a change in the slope of the boiling curve can be observed at point "A" (see Fig. 7). Such a point is called inflection point. After this point, the heat transfer coefficient starts to reduce as can be seen in Fig. 8. In terms of the thermodynamic efficiency, the point "A" is of great importance since it gives a high heat flux with a moderate wall superheat.

**3.3 The effects of the heating surface wettability**

In this section, the influences of the heating surface wettability are studied. Previously it has been mentioned that in the pseudopotential LB model the surface wettability can be adjusted by a fluid-solid interaction force. In this work, we adopt a modified $\psi$-based fluid-solid interaction force [55]:

$$\mathbf{F}_{\text{ads}} = -G_{\text{w}} \psi(\mathbf{x}) \sum_{\alpha} \omega_{\alpha} S(\mathbf{x} + \mathbf{e}_{\alpha}) \mathbf{e}_{\alpha}, \tag{22}$$

where $G_{\text{w}}$ is used to tune the interaction force, $\omega_{\alpha} = w_{\alpha}/3$, and $S(\mathbf{x} + \mathbf{e}_{\alpha}) = \phi(\mathbf{x}) s(\mathbf{x} + \mathbf{e}_{\alpha})$, in which $s(\mathbf{x} + \mathbf{e}_{\alpha})$ is a "switch" function that is equal to 1 or 0 for a solid or a fluid phase, respectively, and $\phi(\mathbf{x})$ can be set to $\psi(\mathbf{x})$. In the present simulations, the modified $\psi$-based fluid-solid interaction is found to give smaller deviations of densities and spurious currents as compared with the $\psi$-based fluid-solid interaction. Two different values are considered for $G_{\text{w}}$: 0.065 and 0.105. With the treatments in the above section, these two choices give contact angles around $50°$ and $55.5°$ (defined in the liquid phase) at $T_s$, respectively. The previous simulations correspond to $G_{\text{w}} = 0$ with the contact angle around $44.5°$.

The density contours obtained with $G_{\text{w}} = 0.065$ at $t = 22000\delta_t$ and $30000\delta_t$ are shown in Figs. 9 and 10, respectively, for the cases $\Delta T = 0.0165$, $0.029$, and $0.032$. Similarly, to make the comparisons between these cases more transparent, the wall superheat is increased at $t = 20000\delta_t$ from an initial wall superheat to the higher wall superheats. From Figs. 9 and 10 we can clearly observe the change from nucleate boiling to film boiling with the increase of the wall superheat. To be specific, it can be seen that the boiling processes at $\Delta T = 0.0165$, $0.029$, and $0.032$ are in the nucleate, transition, and film boiling regimes, respectively, which can be further confirmed by the variations of the transient heat flux in these cases (see Fig. 11).

It is seen that the film boiling occurs at $\Delta T = 0.032$ when $G_{\text{w}} = 0.065$, while the case $\Delta T = 0.032$ shown in the previous section is in the transition boiling regime, which means that, when the surface wettability decreases (the liquid contact angle increases), the film boiling will appear at a lower wall superheat. A similar phenomenon can be observed in Fig. 12 where $G_{\text{w}} = 0.105$. It can be seen that the film boiling occurs in the case $\Delta T = 0.029$ with $G_{\text{w}} = 0.105$. Actually, this is due to the fact that the decrease of the surface wettability will allow the vapor to accumulate at the surface more easily. As pointed out by Xu and Qian [4], if the liquid does not wet the heating surface, the boiling process will immediately enters into the film boiling regime as soon as boiling is initiated.

The boiling curves given by $G_w = 0.065$ and $G_w = 0.105$ are displayed in Fig. 13. For comparison, the previous results ($G_w = 0$) are also shown there. Some information can be obtained from the boiling curves. First, it can be seen that the maximum of the heat flux (the critical heat flux) decreases when the contact angle increases, which confirms the trend revealed by both the experimental and theoretical studies in the literature [56-58], e.g., Chen *et al*. [56] have experimentally found that a Si plain surface (water contact angle $40°$) yields a lower critical heat flux than that given by a SiO$_2$ plain surface (water contact angle $15°$). Furthermore, the aforementioned phenomenon: the boiling process enters into the film boiling regime at a lower wall superheat when the contact angle increases, can also be clearly seen in Fig. 13. This phenomenon has been observed in Takata *et al*.'s experimental study [59]. They found that the Leidenfrost temperature (the minimum surface temperature for film boiling) decreases as the contact angle increases.

Moreover, in Fig. 13 we can observe that the boiling curves cross each other around $\Delta T = 0.015$. After they cross, the heat flux of $G_w = 0$ is larger than that of $G_w = 0.065$, which in turn is larger than the heat flux of $G_w = 0.105$. However, inverse results are seen before their crossing. Such a feature can also be found in some recent experimental studies. For example, a similar crossing of the boiling curves with different contact angles can be seen in Ref. [60]. This feature is due to the different influences of the surface wettability on the critical heat flux and the onset of boiling. With the increase of the contact angle, the critical heat flux will decrease; however, the onset of boiling will be enhanced. Experimental studies have shown [60-62] that the required wall superheat for the onset of boiling will decrease when the contact angle increases. In other words, the onset of boiling occurs earlier on the surfaces with larger contact angles [60], which is also confirmed in our simulations: boiling appears at $\Delta T = 0.01$ when $G_w = 0.105$ (contact angle around $55.5°$) while there is no boiling at the same wall superheat with $G_\text{w} = 0$ (contact angle around $44.5^\circ$). This is the reason why in the early boiling regime the heat flux of $G_\text{w} = 0.105$ is higher than that of $G_\text{w} = 0$.

**4. Summary and conclusion**

In this paper, a hybrid thermal LB model has been developed for simulating thermal multiphase flows, which employs an improved pseudopotential LB model and a finite-difference scheme to simulate the fluid flow and the temperature field, respectively. The pseudopotential LB model and the finite-difference solver of temperature are coupled via the equation of state. Numerical simulations of boiling heat transfer have been conducted using the model. The complete three boiling stages, nucleate boiling, transition boiling, and film boiling, as well as the boiling curve were successfully reproduced in our simulations with the bubble nucleation, growth, departure, and coalescence being well captured. Some basic features of boiling heat transfer were clearly observed in the numerical results, such as the severe fluctuation of transient heat flux in the transition boiling regime, the critical heat flux in the boiling curve, and the feature that the maximum heat transfer coefficient lies at a lower wall superheat than that of the maximum (critical) heat flux.

Furthermore, the influences of the heating surface wettability on boiling heat transfer have also been studied. The numerical results show that the critical heat flux decreases when the contact angle increases, which also makes the boiling process enter into the film boiling regime at a lower wall superheat. Besides, increasing the contact angle will reduce the required wall superheat for the onset of boiling. All of these findings are qualitatively consistent with the experimental studies in the literature, demonstrating that the proposed model is capable of reproducing the basic features and the fundamental characteristics of boiling heat transfer. Future work will be conducted to extend the model to three dimensions and enable the simulation of boiling flows at large density ratios, which may offer quantitative comparisons with experimental results.

**Conflict of interest**

None declared.

**Acknowledgments**

The authors gratefully acknowledge the support from the Los Alamos National Laboratory's Lab Directed Research & Development (LDRD) Program and the Foundation for the Author of National Excellent Doctoral Dissertation of China (No. 201439). This work was performed under the auspices of the National Nuclear Security Administration of the US Department of Energy at Los Alamos National Laboratory under Contract No. DE-AC52- 06NA25396. One of the authors (Q. Li) would like to thank Mr. Anjie Hu very much for helpful discussion.

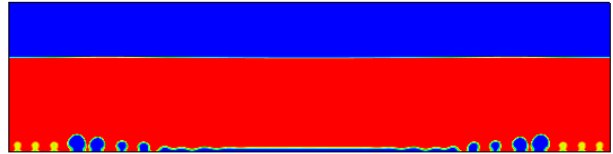

(a) $t = 10000\delta_t$

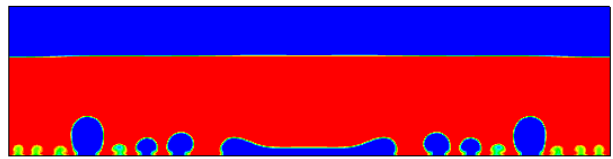

(b) $t = 12000\delta_t$

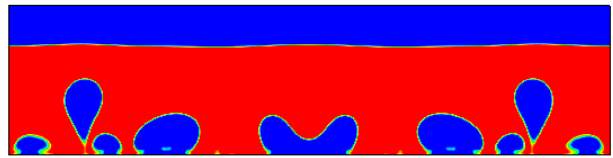

(c) $t = 16000\delta_t$

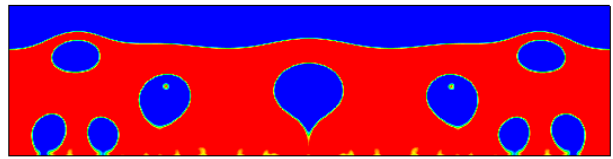

(d) $t = 20000\delta_t$

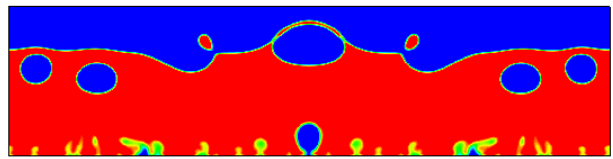

(e) $t = 30000\delta_t$

**Figure 1** Snapshots of the boiling process at $\Delta T = 0.0137$.

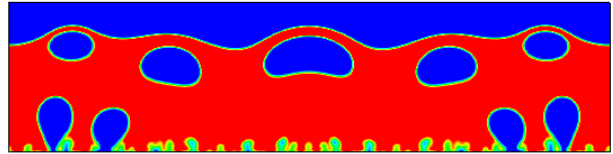

(a) $t = 22000\delta_t$

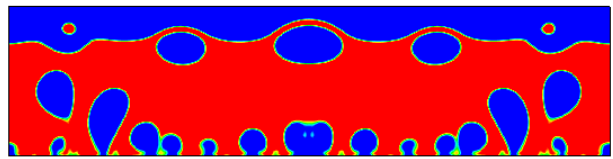

(b) $t = 26000\delta_t$

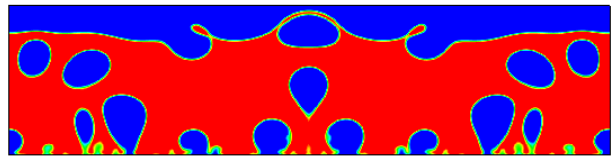

(c) $t = 30000\delta_t$

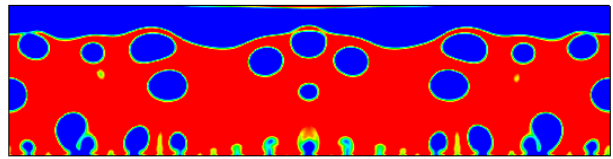

(d) $t = 40000\delta_t$

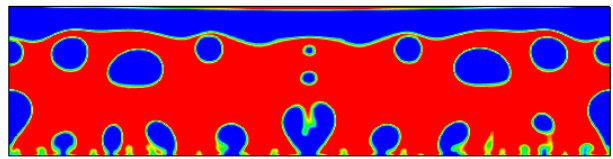

(e) $t = 50000\delta_t$

**Figure 2** Snapshots of the boiling process at $\Delta T = 0.0165$.

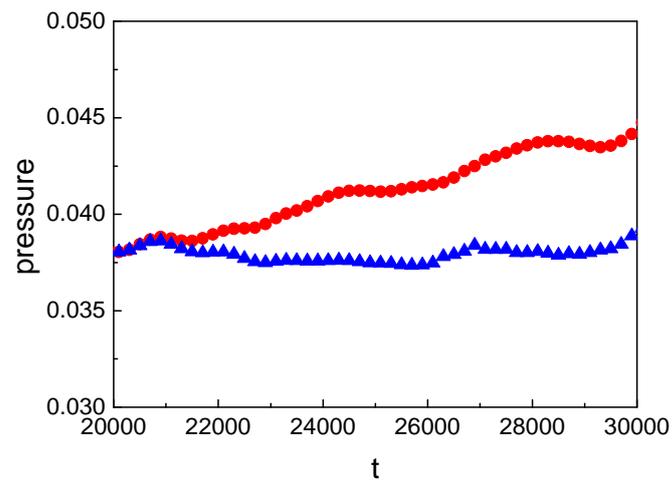

**Figure 3** Variation of the pressure at the central point of the top wall with time. The triangles and the circles represent the results of $\Delta T = 0.0137$ and $0.0165$, respectively.

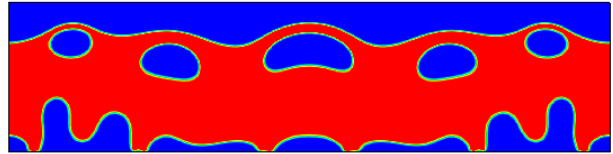

(a)  $t = 22000\delta_t$

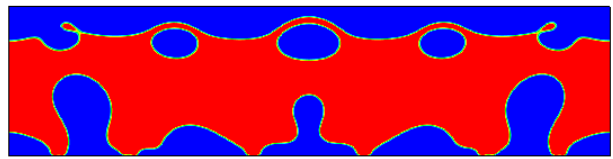

(b)  $t = 26000\delta_t$

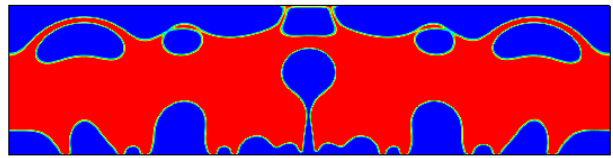

(c)  $t = 30000\delta_t$

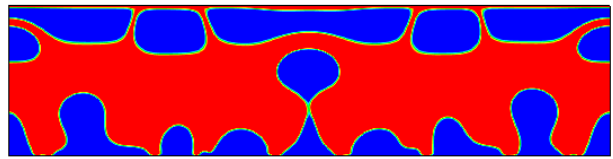

(d)  $t = 40000\delta_t$

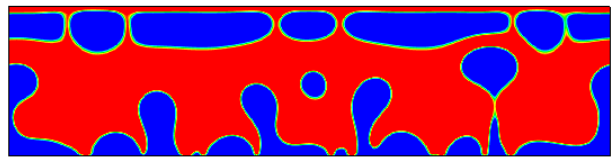

(e)  $t = 50000\delta_t$

**Figure 4** Snapshots of the boiling process at  $\Delta T = 0.032$ .

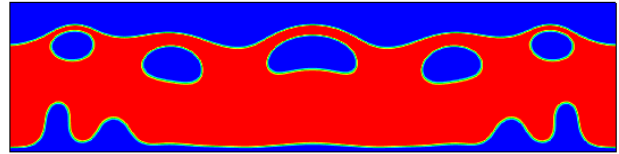

(a) $t = 22000\delta_t$

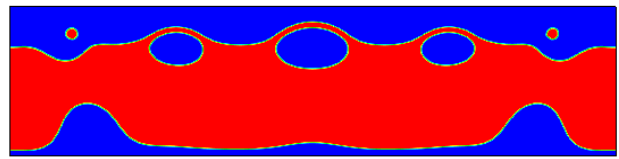

(b) $t = 26000\delta_t$

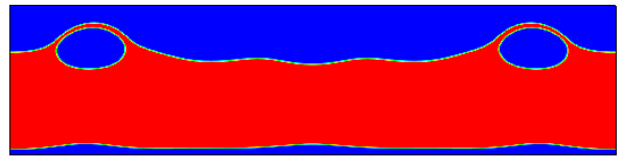

(c) $t = 35000\delta_t$

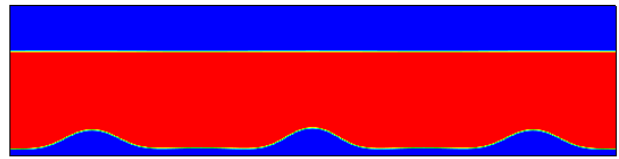

(d) $t = 120000\delta_t$

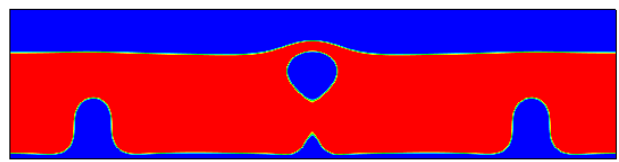

(e) $t = 140000\delta_t$

**Figure 5** Snapshots of the boiling process at $\Delta T = 0.035$.

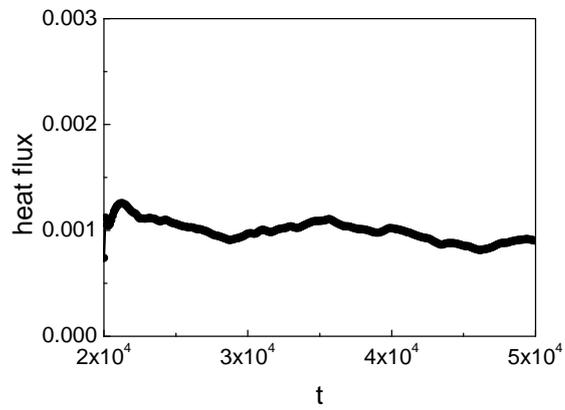

(a) $\Delta T = 0.0165$

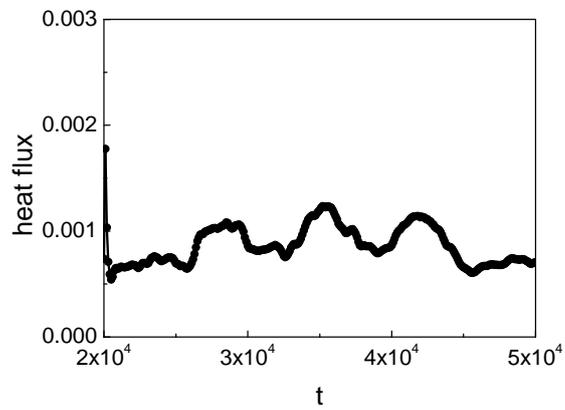

(b) $\Delta T = 0.032$

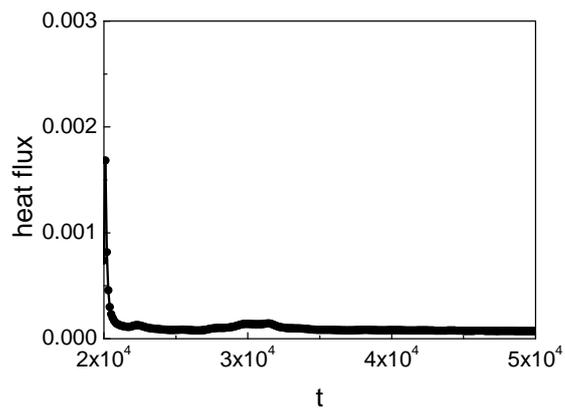

(c) $\Delta T = 0.035$

**Figure 6** Variations of the transient heat flux $q(t)$ in the cases $\Delta T = 0.0165$, $0.032$, and $0.035$.

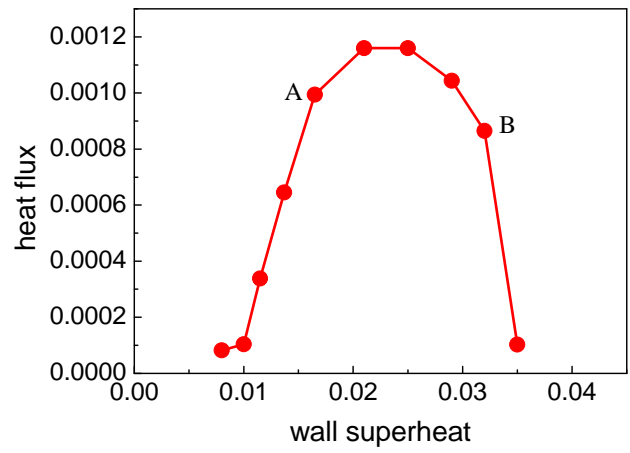

**Figure 7** The predicted boiling curve (the average heat flux against the wall superheat). The points "A" and "B" denote the cases $\Delta T = 0.0165$ and $0.032$, respectively.

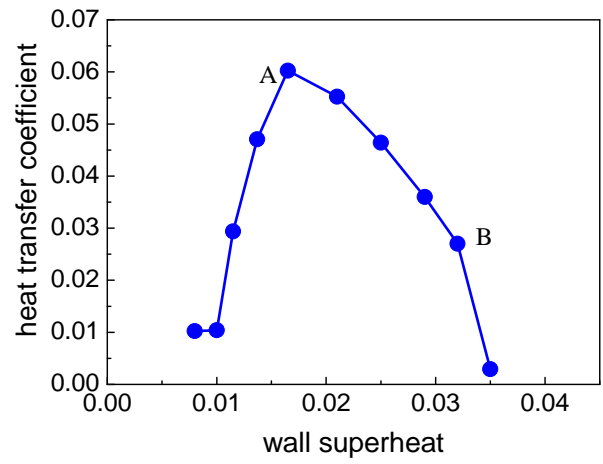

**Figure 8** Variation of the heat transfer coefficient $h_c = \bar{q}/\Delta T$. The points "A" and "B" denote the cases $\Delta T = 0.0165$ and $0.032$, respectively.

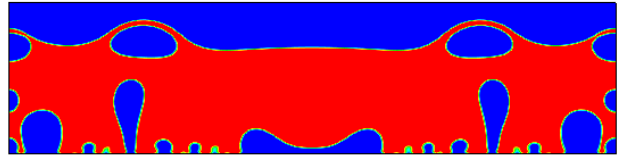

(a) $\Delta T = 0.0165$

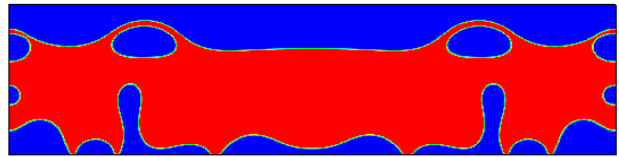

(b) $\Delta T = 0.029$

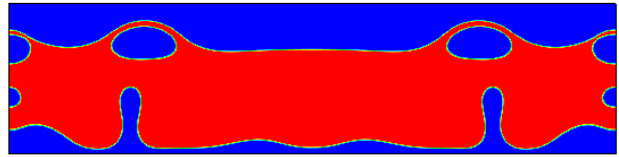

(c) $\Delta T = 0.032$

**Figure 9** Snapshots of boiling processes at $t = 22000\delta_t$ for the cases $\Delta T = 0.0165$, $0.029$, and $0.032$ with $G_\mathrm{w} = 0.065$.

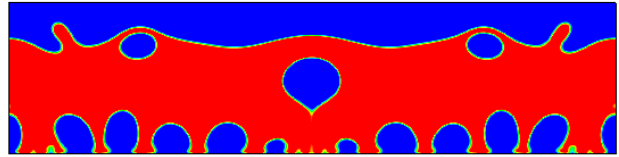

(a) $\Delta T = 0.0165$

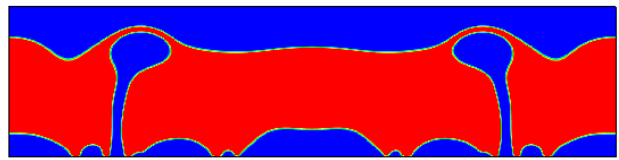

(b) $\Delta T = 0.029$

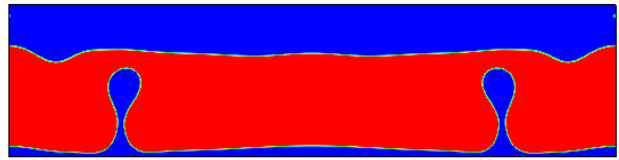

(c) $\Delta T = 0.032$

**Figure 10** Snapshots of boiling processes at $t = 30000\delta_t$ for the cases $\Delta T = 0.0165$, $0.029$, and $0.032$ with $G_\mathrm{w} = 0.065$.

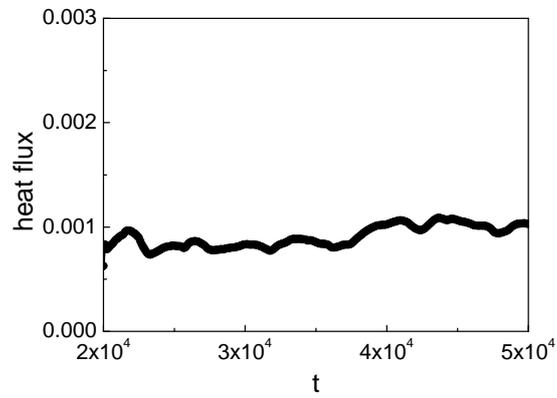

(a) $\Delta T = 0.0165$

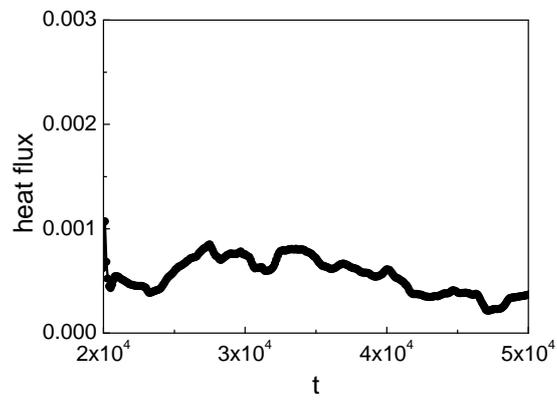

(b) $\Delta T = 0.029$

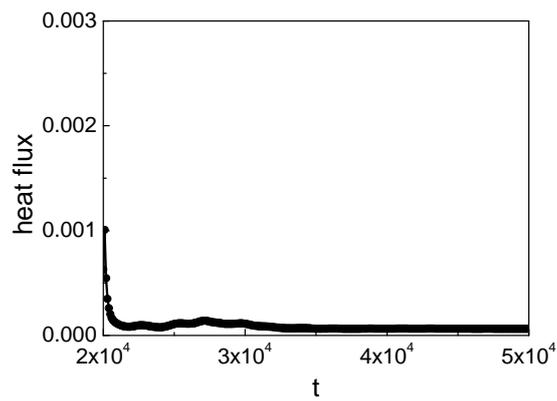

(c) $\Delta T = 0.032$

**Figure 11** Variations of the transient heat flux in the cases $\Delta T = 0.0165$, $0.029$, and $0.032$ with $G_\mathrm{w} = 0.065$.

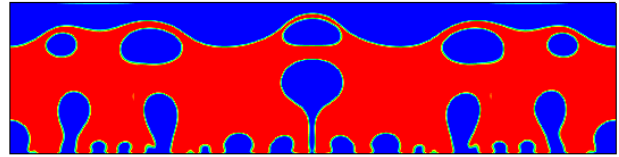

(a) $\Delta T = 0.0165$

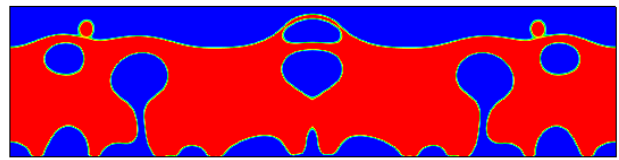

(b) $\Delta T = 0.025$

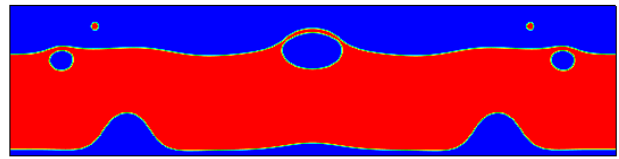

(c) $\Delta T = 0.029$

**Figure 12** Snapshots of boiling processes at $t = 30000\delta_t$ for the cases $\Delta T = 0.0165$, $0.025$, and $0.029$ with $G_w = 0.105$.

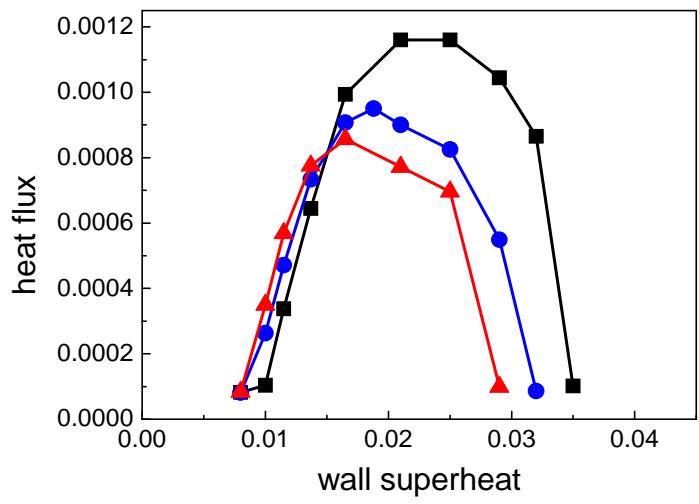

**Figure 13** Comparison of the boiling curves. The squares, circles, and triangles represent the results of $G_\mathrm{w} = 0$, $0.065$, and $0.105$, respectively, with the corresponding static (liquid) contact angles around $44.5°$, $50°$, and $55.5°$.